\documentclass{article}

\usepackage{arxiv}

\usepackage[utf8]{inputenc} 
\usepackage[T1]{fontenc}    
\usepackage{hyperref}       
\usepackage{url}            
\usepackage{booktabs}       
\usepackage{amsfonts}       
\usepackage{nicefrac}       
\usepackage{microtype}      
\usepackage{lipsum}		
\usepackage{graphicx}
\usepackage{doi}
\usepackage[square,sort,comma,numbers]{natbib}

\title{LEBS Toolkit for Addressing Development Challenges through Astronomy}

\author{ \href{https://orcid.org/0000-0002-9745-0504}{\includegraphics[scale=0.06]{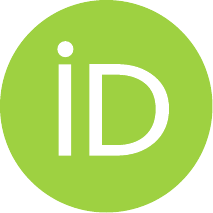}\hspace{1mm}Joyful E. Mdhluli}{ on behalf of the IAU Office of Astronomy for Development} \thanks{Visit our website, www.astro4dev.org} \\
	International Astronomical Union's Office of Astronomy for Development\\
	South African Astronomical Observatory\\
        Cape Town, South Africa\\
	\texttt{joy@astro4dev.org} \\
	\And
}

\renewcommand{\shorttitle}{\it{LEBS Toolkit}}

\hypersetup{
pdftitle={LEBS Toolkit for Addressing Development Challenges through Astronomy},
pdfsubject={astro-ph.IM},
pdfauthor={Joyful E.~Mdhluli},
pdfkeywords={Astronomy, Sustainable development, Toolkit},
}

\begin{document}
\maketitle

\begin{abstract}
The LEBS Toolkit provides a comprehensive framework for astronomers and natural scientists to address development challenges through their field. The toolkit aligns with the Sustainable Development Goals (SDGs) and offers practical actions to achieve development outcomes. It emphasises the importance of effective communication between astronomy and development, highlighting four key components: Leadership, Externalities, Break Barriers to Entry and Support Functioning Markets \& Institutions (LEBS). Using these components, scientists can promote accurate information dissemination, encourage positive community impacts, overcome monopolistic barriers, and improve market efficiency and institutional settings. The LEBS Toolkit serves as a guide for the scientific community in bridging the gap between scientific knowledge and practical development actions, ultimately contributing to global development goals.
\end{abstract}


\section{Introduction}
In an ever-evolving world, the intersection of science and societal development holds profound potential. The United Nations Sustainable Development Goals\href{https://sdgs.un.org/goals}{(SDGs)}\footnote{https://sdgs.un.org/goals} serve as a global benchmark to measure progress towards a better life for all. However, the connection between blue-sky sciences such as astronomy and the attainment of these goals is not always immediately apparent. This gap presents a unique opportunity for astronomers and natural scientists to contribute to development in meaningful ways. The Leadership, Externalities, Break Barriers to Entry and Support Functioning Market (LEBS) Toolkit, developed by Dr. Tawanda Chingozha, a past fellow at the International Astronomical Union's Office of Astronomy for Development (IAU OAD) and Development Economist, is designed to bridge this gap by providing practical guidance on how scientific endeavours can address development challenges. Based on economic principles, the toolkit emphasises the role of leadership, externalities, barriers to entry, and functioning markets and institutions in achieving sustainable development. Using the expertise and knowledge of the scientific community, the LEBS Toolkit aims to transform theoretical understanding into actionable steps, ultimately fostering positive change and development outcomes.

\section{Conceptual Framework}
The Sustainable Development Goals (SDGs) are a universal benchmark for conceptualising and measuring the various aspects of development. Each SDG has a set of targets that further unpack the various interventions that communities and governments may work on to achieve a better life for all.

\begin{figure}
    \centering
    \includegraphics[width=0.6\textwidth]{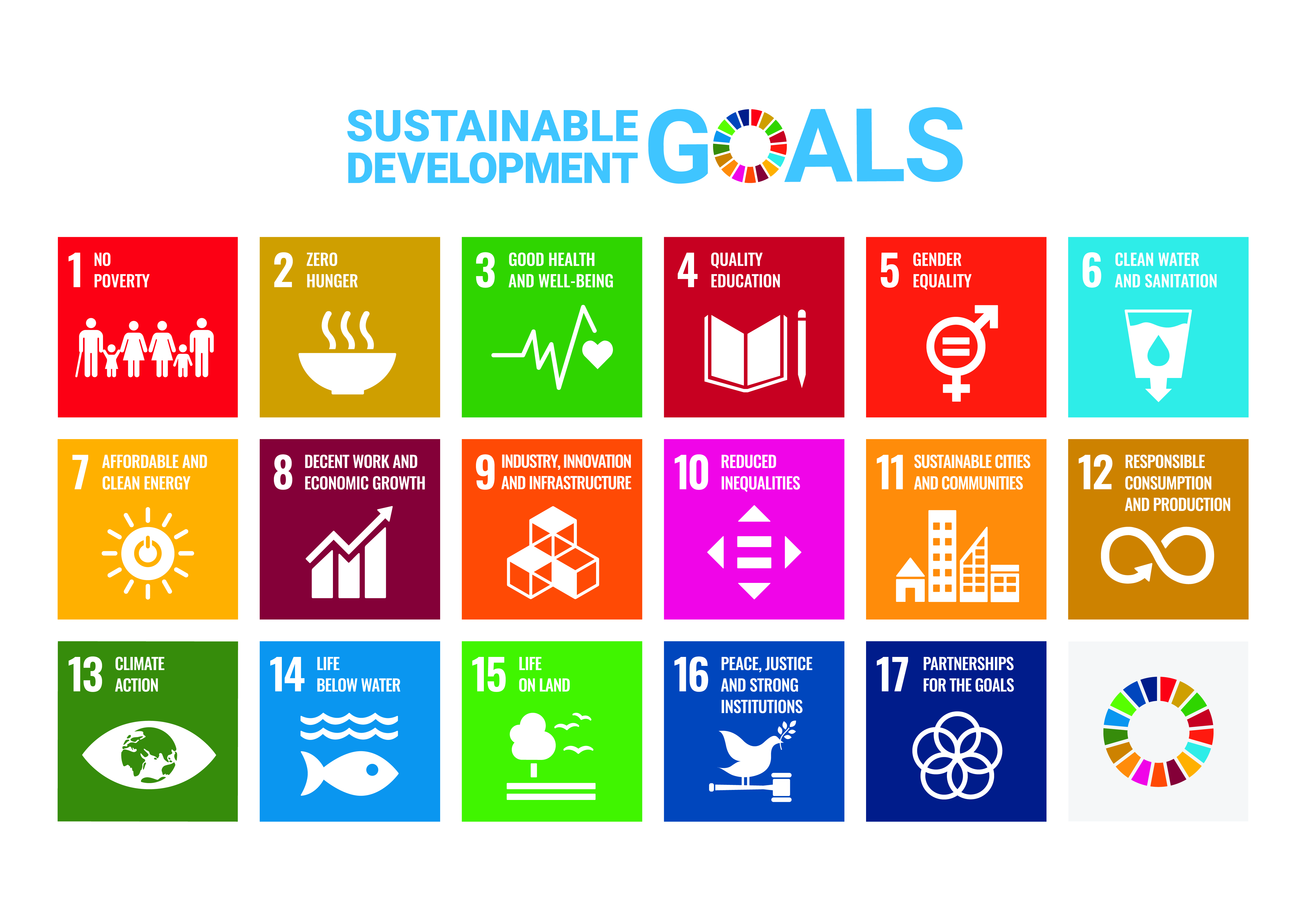}
    \caption{United Nations Sustainable Development Goals. Image Credit: \href{https://sdgs.un.org/goals}{United Nations Sustainable Development Goals}}
    \label{fig:enter-label}
\end{figure}

The SDGs are concise and clear-cut, but this may not necessarily hold true when one thinks about how blue-sky sciences such as astronomy can contribute to the attainment of the SDGs. It is one thing to understand the SDGs as a natural scientist, and it may be another to know some of the day-to-day activities that one can do towards achieving different development outcomes or SDGs. This is the gap that the LEBS Toolkit seeks to plug. The toolkit provides guidance on some day-to-day actions natural scientist can do to make a development impact; where impact is measured using the SDGs and/or SDG targets. The LEBS Toolkit is a Call to Action whose main points are derived from Economics.

\section{LEBS Overview}
The toolkit is built on four core pillars: \textbf{L}eadership, \textbf{E}xternalities, \textbf{B}reak Barriers to Entry, \textbf{S}upport Functioning Markets \& Institutions. At the center of the LEBS Toolkit is the assumption that many development challenges faced by the world emanate from market failure. Market failure is an economic phenomenon in which the market forces of demand and supply fail to allocate resources efficiently. There are many causes of market failure, and here we discuss only those causes that are related to the 4 components of the LEBS Toolkit as follows:

\begin{itemize}
    \item \textbf{Leadership}\\
    It is difficult to control the flow of information. Markets on their own cannot limit the amount of false/toxic information (including baseless conspiracies) that people have access to. Someone somewhere will always have the incentive to spread false or even harmful information. Scientists/astronomers may have a leadership role in helping circulate the right information within their influence circles. A good example is vaccine hesitancy that dominated headlines during the COVID-19 pandemic. Scientists/astronomers may play an important leadership role to help correct inaccurate views/beliefs regarding different matters/issues that affect society (including vaccine hesitancy).
    \item \textbf{Externalities}\\
    An economic externality is a phenomenon where a transaction/activity affects a third person/entity that is not involved. An example of a negative externality is when a fishing community is affected by water pollution that is caused by a factory that discharges effluent in the river upstream. A positive externality occurs when for instance a mining company constructs a dam to meet their water requirement, but also allow the surrounding community to access the water resource. Scientists/astronomers may allow surrounding communities to benefit from their activities/resources (even though they were not necessarily earmarked for the communities) as a manifestation of positive externalities. At the same time, scientists/astronomers may use their skills/knowhow in instrumentation to produce sensors that measure noise (or other forms of pollution) and assist authorities take corrective measures – thereby minimising the occurrence and impact of negative externalities.
    \item \textbf{Break Barriers to Entry}\\
    Monopolies have plenty of leverage in determining the selling price of their goods and services, and resultant higher prices may exclude whole communities. Monopolies in medicine, medical equipment and other sectors exist due to barriers to entry. Barriers may include government restrictions (say for example to establish a private nuclear station), high start-up costs, strategic national importance (for water and electricity utilities in some countries) or just because there is insufficient knowledge. Where monopolies for certain goods that can help improve people’s lives exist because of lack of information, scientists have an important role to play since they are usually the custodians of information regarding different scientific breakthroughs. At the height of the COVID-19 pandemic, astronomers and engineers in South Africa responded to shortages of ventilators by producing prototype ventilators -  hence breaking down barriers to entry.
    \item \textbf{Support Functioning Markets \& Institutions}\\
    Development challenges thrive in environments where markets do not efficiently work (transport, employment etc.) and also where the institutional setting is poor (harmful stigmas \& taboos, corruption and lack of transparency etc.). Through application programming and machine learning for example, astronomers/scientists can more effectively match job seekers to jobs, or improve food distribution so that some do not go hungry while others throw away food in a particular area. In poor institutional settings, the scientific community may take a lead in rejecting regressive tendencies such as discrimination and harmful taboos. E-governance (a product of application programming) may also help promote less human interaction and opportunities for malfeasance in situations where corruption and the lack of transparency are pervasive. 
\end{itemize}

\begin{figure}[h!]
    \centering
    \includegraphics[width=0.6\textwidth]{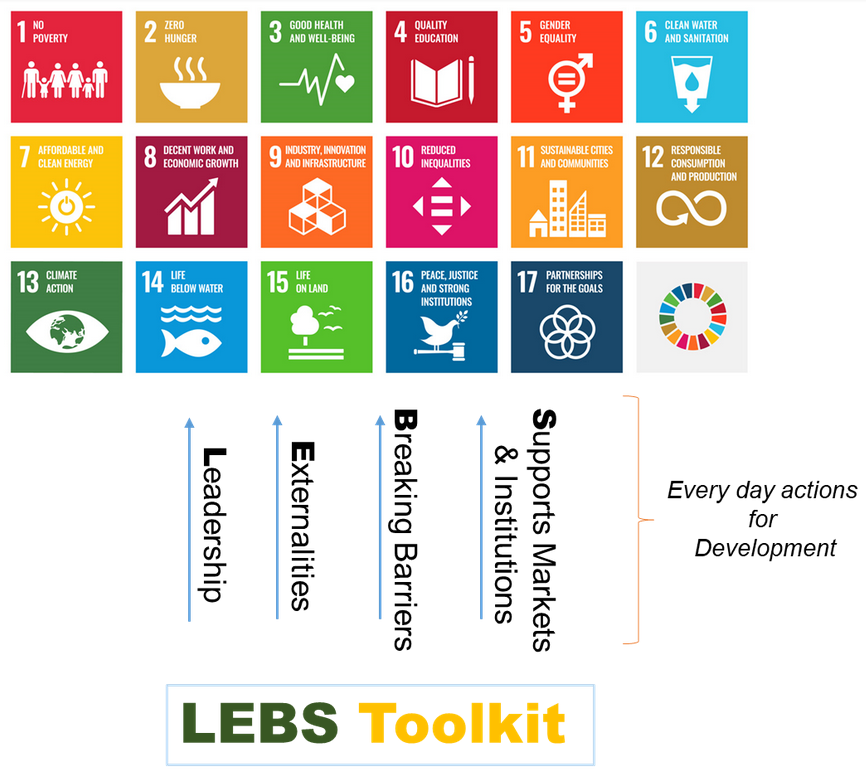}
    \caption{The LEBS Toolkit and SDGs}
    \label{Toolkit}
\end{figure}

\begin{figure}
    \centering
    \includegraphics[width=0.2\textwidth]{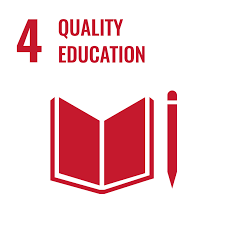}
    \includegraphics[width=0.2\textwidth]{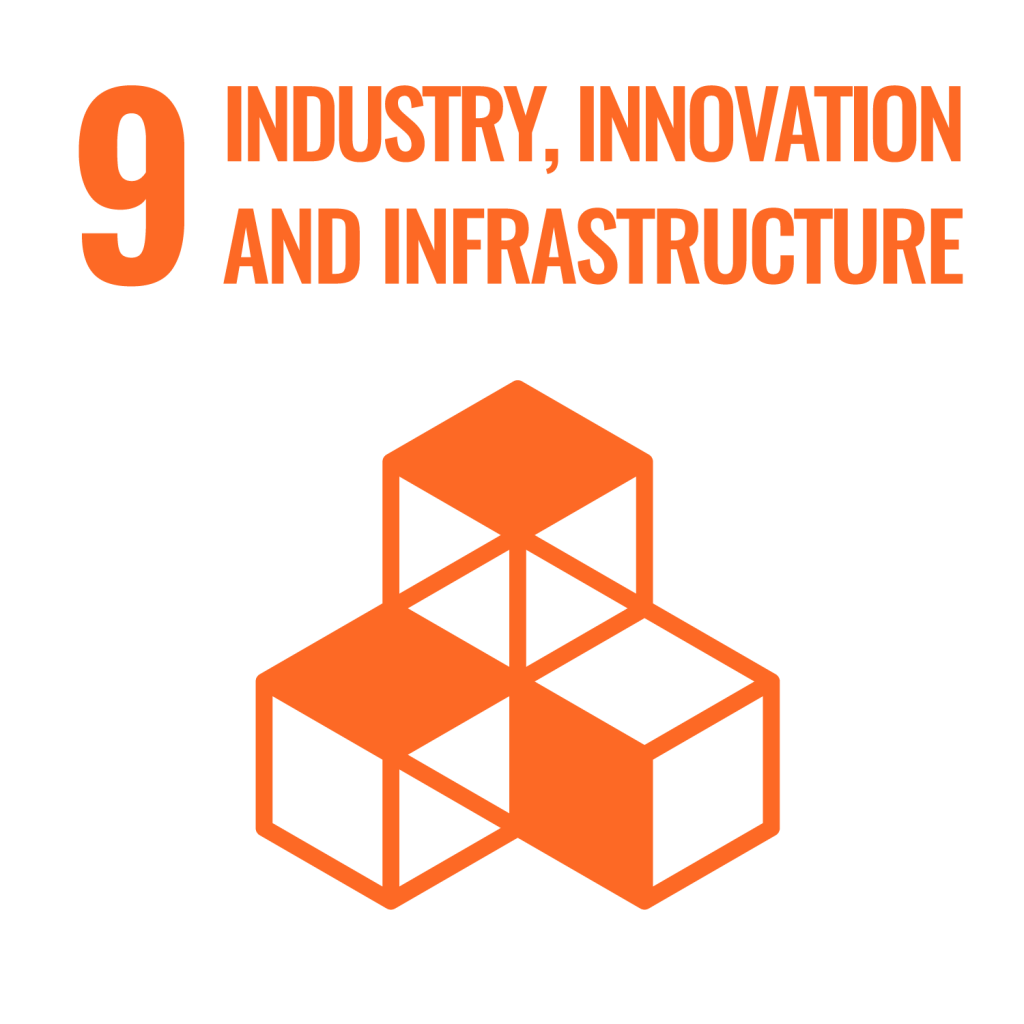}
    \includegraphics[width=0.2\textwidth]{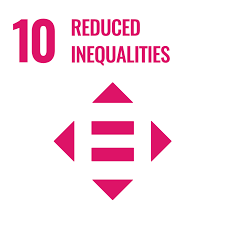}
    \caption{Examples of Sustainable Development Goals (SDGs) that converge with with the toolkit. Image Credit: \href{https://sdgs.un.org/goals}{United Nations Sustainable Development Goals}}\label{sdgs}
\end{figure}

\textbf{Alignment with the UN SDGs}\\
The toolkit directly contributes to several SDGs (see Fig. \ref{Toolkit}), particularly in promoting quality education, reducing inequalities, and fostering innovation and infrastructure as shown in Fig. \ref{sdgs}. By integrating astronomy with economic principles, it provides a pathway for interdisciplinary collaboration and sustainable progress.

\section{Why LEBS Toolkit is important for communicating astronomy}
Astronomers often face the challenge of demonstrating the tangible benefits of their field to others. Despite significant work on the economic value of astronomy, the connection between astronomy and development remains difficult to make for many. The Office of Astronomy for Development (OAD) has been at the forefront of this effort \cite{comment1}-\cite{comment2}, identifying various thematic areas and flagship projects that translate astronomical advancements into development outcomes such as astronomy for socio-economic development \cite{resources}, astronomy for mental wellbeing and astronomy skills and knowledge for development. The LEBS Toolkit builds upon the Sustainable Development Goals (SDGs) and the OAD's existing work, particularly its flagship projects, to illuminate the key actions and channels through which astronomers can contribute to development. By providing a practical framework, the toolkit assists the astronomy community in effectively communicating the link between their scientific endeavours and broader development goals. This approach not only highlights the relevance of astronomy but also showcases its potential to drive positive change and development

\subsection{Applications in Development}
Several practical applications of the LEBS Toolkit include:
\begin{itemize}
\item Utilising astronomy education to bridge knowledge gaps in underserved communities \cite{chapman}.
\item Leveraging astronomical data for technological and infrastructure advancements \cite{govender}.
\item Encouraging policies that enhance scientific inclusivity and economic participation.
\end{itemize}

\section{Application of the Toolkit}
The LEBS Toolkit encourages scientists and astronomers to take actionable steps in addressing development challenges. A worksheet is provided to help users think practically about these actions. Users are prompted to select one or two development challenges from a list of examples provided below or one of their own choosing and identify which LEBS components - Leadership, Externalities, Break Barriers to Entry, or Support Functioning Markets \& Institutions - are best suited to tackle each challenge. Users are also encouraged to justify their choices with brief remarks in the worksheet's designated column, see Table \ref{tab:worksheet}. This approach ensures that scientists can systematically apply the toolkit to real-world problems and contribute effectively to sustainable development goals.

{\renewcommand{\arraystretch}{2}%
\begin{table}[h]
    \centering
    \begin{tabular}{|c|c|c|}
        \hline
        \textbf{LEBS Toolkit Component} & \textbf{Tick} & \textbf{Justify/Remarks} \\
        \hline\hline
        Leadership &  &  \\
        \hline
        Externalities &  & \\
        \hline
        Break Barriers to Entry &  & \\
        \hline
        Support Functioning Markets \& Institutions &  & \\
        \hline
        (Other Component) &  & \\
        \hline
    \end{tabular}
    \caption{Template of the LEBS Toolkit Worksheet}
    \label{tab:worksheet}
\end{table}

\textbf{Examples of Development Challenges}\\
There is a variety of examples that are classified as development challenges, including poverty, hunger, conflict, climate change, racism, discrimination, unemployment, and inequality. The IAU Office of Astronomy for Development (OAD) has funded a variety of projects that address these challenges, and a few examples are listed below (more details of the projects are available on the OAD website \cite{casestudies}).

In Chile, the “Rediscovering Identity Through Astronomy” project targeted indigenous youth in remote communities with limited access to STEM education. By engaging students from the Mapuche, Likan Antai, and Pascuense communities in monthly astronomy-themed questions followed by virtual discussions, the project enhanced both scientific literacy and cultural awareness. Participants increased their engagement and confidence, rediscovered ancestral stories, and built cross-community connections, with some expressing interest in pursuing astronomy professionally. This project demonstrates how astronomy can simultaneously promote science education and cultural identity, addressing inequality in educational access and supporting sustainable development goals (SDGs) 4 (Quality Education) and 11 (Sustainable Cities and Communities).

In Bangladesh, an interactive mobile application was developed to improve astronomy education for students in rural and underserved areas. The app used anthropomorphized planetary and lunar characters to teach astronomy concepts alongside environmental topics such as climate change and light pollution. In pilot testing with seventy-nine students, approximately one-third showed measurable learning gains, while nearly all participants reported enjoying the experience and gaining new knowledge. This initiative illustrates the potential of digital platforms to deliver engaging, accessible science education to marginalized communities, contributing to SDG 4 (Quality Education).

The potential of astronomy to support mental health has been explored through workshops conducted in Armenia, Spain, and South Africa. These workshops combined astronomy with cognitive, emotional, and social support frameworks, targeting children with mental-health needs, elderly care-home residents, and professional caregivers. Participants reported increased motivation, improved mood, and enhanced self-esteem, highlighting astronomy’s capacity to inspire awe, promote social interaction, and provide psychosocial benefits. This work addresses challenges related to conflict, stress, and well-being, aligning with SDG 3 (Good Health and Well-Being).

In Africa, Big Data hackathons have used astronomy’s data-rich environment to build technical and workforce skills. Students participated in lectures, workshops, and three-day hackathons using real datasets to solve practical problems, with approximately 400 students trained and many moving into data-science careers. The proportion of female participants increased from 20\% to nearly 50\%, demonstrating the project’s contribution to gender equity. These initiatives address challenges related to unemployment and inequality while supporting SDGs 4 (Quality Education), 8 (Decent Work and Economic Growth), 10 (Reduced Inequalities), and 17 (Partnerships for the Goals).

In Pakistan, high-quality, culturally relevant astronomy content was produced for schoolchildren in Urdu through video lessons aligned with the national curriculum. By integrating local cultural references and bilingual subtitles, the lessons reached thousands of students via television broadcasts and online platforms, increasing access to science education in underserved areas. This initiative helps address inequality and supports SDGs (Quality Education), 10 (Reduced Inequalities), 16 (Peace, Justice and Strong Institutions), and 17 (Partnerships for the Goals).

Finally, the OruMbya project in Brazil combined astronomy with Afro-Brazilian and Indigenous knowledge to promote cultural identity, social inclusion, and empowerment among marginalized youth. Through webinars and the creation of a symbolic “sky-inspired” garden, the project engaged approximately 1,400 participants, who reported enhanced cultural understanding and community cohesion. This work demonstrates how astronomy can bridge science with cultural heritage and social development, addressing challenges such as discrimination and inequality while supporting SDGs 4 (Quality Education), 10 (Reduced Inequalities), 11 (Sustainable Cities and Communities), 13 (Climate Action), and 17 (Partnerships for the Goals).

Collectively, these examples illustrate the versatility of astronomy as a tool for addressing development challenges. When tailored to local contexts and integrated with cultural, educational, or social objectives, astronomy projects can foster equitable access to education, support mental well-being, promote technical skills, and enhance cultural understanding, thereby contributing to multiple dimensions of sustainable development  \cite{resource4}-\cite{comment4}.

\section{Conclusion}
In conclusion, the LEBS Toolkit offers astronomers and natural scientists a practical framework to address development challenges through the application of their expertise. By aligning with the Sustainable Development Goals (SDGs) and focusing on key components such as Leadership, Externalities, Breaking Barriers to Entry, and Supporting Functioning Markets and Institutions, the toolkit empowers scientists to make a tangible impact on development outcomes. Through systematic application and thoughtful action, the scientific community can bridge the gap between theoretical knowledge and real-world development, ultimately contributing to a more sustainable and equitable future. The LEBS Toolkit not only highlights the importance of scientific contributions to development but also provides a clear roadmap for achieving meaningful and lasting change.

\section*{Acknowledgments}
The IAU Office of Astronomy for Development acknowledges Dr. Tawanda Chingozha for developing this toolkit.

\end{document}